\documentclass{iaus}
\usepackage{graphicx}

\title[Maser emission in PNe] 
{Maser emission in planetary nebulae}

\author[G\'omez, Y.]   
{Yolanda G\'omez$^1$%
}
\affiliation{$^1$Centro de Radioastronom\'\i a y Astrof\'\i sica, UNAM, 
A.P. 3-72, C.P. 58090, M\'exico  \break email: y.gomez@astrosmo.unam.mx\\[\affilskip]
}
\pubyear{2007}
\volume{xxx}  
\pagerange{119--126}
\date{?? and in revised form ??}
\jname{Proceedings Title IAU Symposium}
\editors{A.C. Editor, B.D. Editor \& C.E. Editor, eds.}
\begin{document}

\maketitle

\begin{abstract}
Stars at the top of the asymptotic giant branch (AGB) can exhibit maser
emission from molecules like SiO, H$_2$O and OH. These masers appear
in general stratified in the envelope, with the SiO masers close to the
central star and the OH masers farther out in the envelope. As the star
evolves to the planetary nebula (PN) phase, mass-loss stops and ionization
of the envelope begins, making the masers disappear progressively. The OH 
masers in PNe can be present in the envelope for periods of 
$\sim$1000 years but the H$_2$O 
masers can survive only hundreds of years. 
Then, H$_2$O maser emission is not
expected in PNe and its detection suggests that these objects are
in a very particular moment of its evolution in the transition from AGB to
PNe. We discuss the unambiguous detection of H$_2$O 
maser emission in two planetary nebulae: K~3-35 and IRAS~17347-3139.
The water-vapor masers in these PNe are tracing disk-like structures around
the core and in the case of K3-35 the  masers were also
found at the tip of its bipolar lobes. Kinematic modeling of the H$_2$O 
masers in both PNe suggest the existence of a rotating and expanding disk.
Both PNe exhibit a bipolar morphology and in the particular
case of K~3-35 the OH masers are highly polarized close to the core in a
disk-like structure. All these observational results are consistent with the models 
where rotation and magnetic fields have been proposed to explain the asymmetries
observed in planetary nebulae.  

\end{abstract}

\firstsection 
\section{Introduction}

Planetary nebulae (PNe) represent the final stage of evolution for intermediate
mass stars (1-8 M$_\odot$). A PN consists of an expanding gaseous shell of
highly ionized gas that has been ejected by the star
during the end of the AGB evolution, and a central star that will end as a 
white dwarf.  The study of PN is very important not only because
our Sun will end as a PN but also because these objects expel heavy 
elements like carbon and oxygen producing an enrichment of the interstellar 
medium. The lifetime of a PNe is about 10$^4$ years and it is expected that 
there are 10$^4$-10$^5$ PNe in our galaxy, from which only $\sim$1500 have 
been reported (\cite{Kohoutek01}). PNe are characterized by exhibiting a 
variety of shapes (e.g. \cite{Balick02} ). 
Several observational studies 
have revealed that most of the
PNe ($\sim$75$\%$; \cite{Manch04}) exhibit asymmetric morphologies that go from 
elliptical to very collimated outflows. One of the open questions in this
field is to understand the origin of these asymmetries.

\begin{table}\def~{\hphantom{0}}
  \begin{center}
  \caption{OH and H$_2$O maser emission in young PNe}
  \label{tab:young}
   \begin{tabular}{lcccc}\hline
      PNe          & OH   & H$_2$O   & Morphology & References \\\hline
     NGC~6302    & YES  & NO        & bipolar & Payne et al. (1988)\\
     OH~349.36-0.20 &YES&NO&?& Zijlstra et al. (1989)\\
     IRAS 17207-2855& YES & NO  & ? & Zijlstra et al. (1989) \\
     PK 356+2$^\circ$ 1 & YES & NO  & ? & Zijlstra et al. (1989) \\
     IRAS 17347-3139 & YES & YES  & bipolar & de Gregorio-Monsalvo et al. (2004) \\
     IRAS 17371-2747 & YES & NO &  ? & Zijlstra et al. (1989) \\
     IRAS 17375-2759 & YES & NO &  ? & Zijlstra et al. (1989) \\
     IRAS 17375-3000 & YES & NO &  ? & Zijlstra et al. (1989) \\
     OH0.9+1.3 & YES & NO & ? & Zijlstra et al. (1989) \\
     IRAS 17443-2949 & YES & YES &  ? & Zijlstra et al. (1989); Su\'arez et al. (2007)\\
     IRAS 17580-3111 & YES & YES$^a$ &  ? &  Zijlstra et al. (1989); Su\'arez et al. (2007)\\
     IRAS 18061-2505 & NO  & YES$^b$  & ? & Su\'arez et al. (2007)\\
     IC~4997         & YES & NO       & bipolar & \cite{Tamura89} \\
     K~3-35          & YES & YES & bipolar & Engels et al. (1989); Miranda et al. (2001)\\
     Vy2-2  &YES&NO&shell& \cite{Davis79} \\
     M1-92 & YES & NO & bipolar & \cite{lepine74} \\
\hline
\end{tabular}
\begin{tabular}{l}
\noindent \scriptsize{$^a$ Probably not a PN; $^b$ No OH maser emission (\cite{Suarez07}).}\\
\end{tabular}
 \end{center}
\end{table}

To answer this question we should review the formation process itself in the
transition between the AGB and PNe stages. After completion of the H and He core 
burning, the star evolves to the AGB.
At the top of the AGB phase, the star is surrounded by an expanding envelope
of gas and dust (\cite{Casw74} ). 
The study of the star in this evolutionary phase is better pursued with radio 
and infrared telescopes and the presence of neutral atomic and molecular gas
in the envelope of the star is common (e.g 
\cite{Mufson75}; \cite{Huggins96}; \cite{Josselin03}; \cite{Bujarrabal06} ). 
For AGB stars with oxigen-rich envelopes it is possible to detect 
maser emission of one or more molecules, such as SiO, H$_2$O and OH that 
appear stratified in the envelope (e.g. \cite{reidmoran81}; \cite{Elitzur92} ). 
As the slow and massive mass loss of the late AGB phase stops, the gravity contracts 
the core heating it up, and the star enters its PN phase, 
from this moment the maser conditions 
will disappear in short timescales (\cite{Lewis89} , \cite{Gomez90}). In particular, the water 
molecules are expected to 
disappear in a timescale of decades, and only OH masers seem to persist for a 
considerable time ($\sim$1000 yr; \cite{Kwok93}). In this way H$_2$O masers are not
expected to be present in a regular PN. However, two PNe 
(K3-35 and IRAS 17347-3139; \cite{Miranda01}; \cite{deGre04} ) 
have been found to harbor H$_2$O and OH maser emission, suggesting that 
these objects can be in an early stage of their evolution as PNe. Recently, two
more PNe have been reported to exhibit H$_2$O maser emission and interferometric
observations are required to confirm the association (\cite{Suarez07}). 
In this talk we will present details about the kinematics and 
polarization of the maser emission in the two confirmed PNe: K3-35 and 
IRAS~17347-3139.  

\begin{figure}
\includegraphics[height=5.3in,width=5.3in,angle=0]{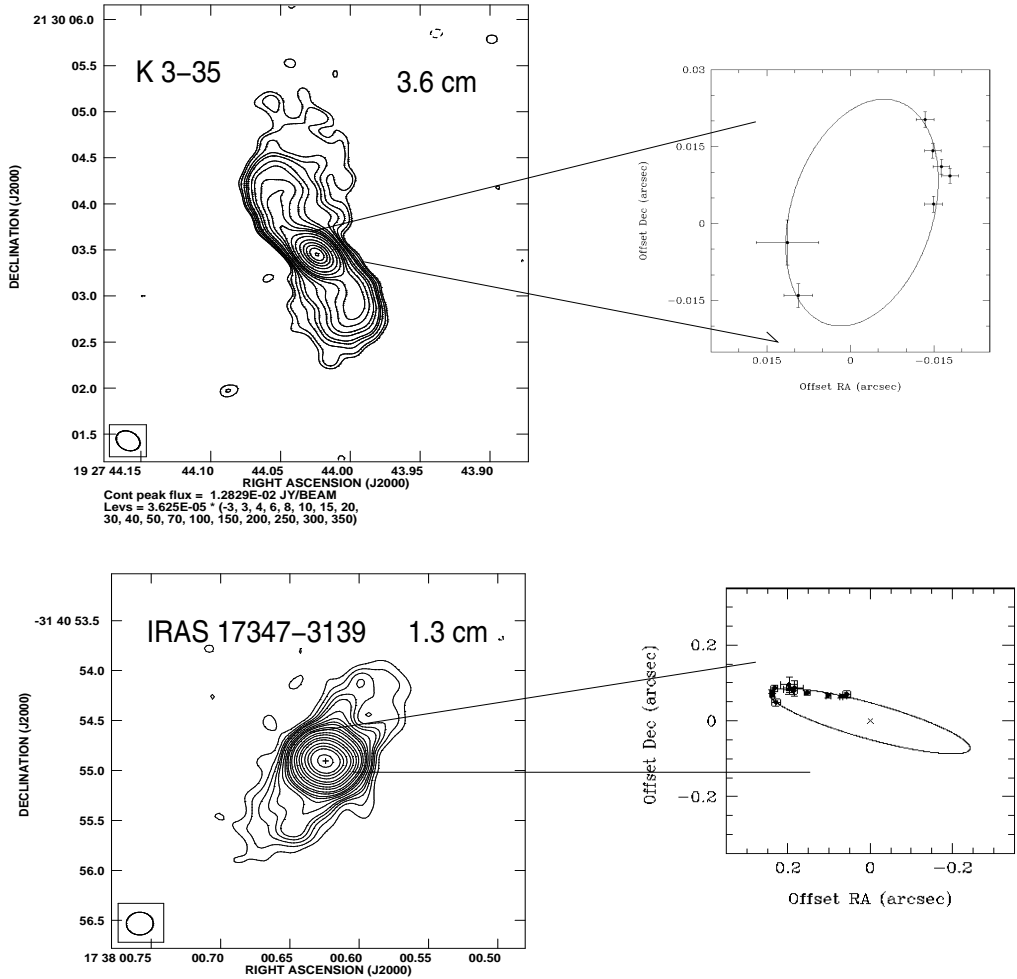}
  \caption{Left: Contour images of the radio continuum observed with the VLA
in K~3-35 (\cite{Miranda01}) and IRAS~1737-3139 (\cite{Tafoya07b}). Right: The
respective water maser spots with a disk fit (\cite{Uscanga07}).
}\label{fig:torus}
\end{figure}

\section{Young PNe with OH and H$_2$O masers}\label{sec:Ohmasers}

Modeling of the OH maser emission in late AGB stars has shown 
that the typically observed double peak profile corresponds, in general,
to the blueshifted and redshifted maser components which are coming mainly 
from the front and the back parts of the envelope, respectively (\cite{reid77}).  
We mentioned that after the mass loss stops the OH maser can survive around 
$\sim$10$^3$ years, then it may be possible to find OH maser emission in very 
young PNe. In the particular case when an ionized region is present, which is 
the case for young PNe, the redshifted OH maser peak component coming from the back,
may not be seen because the emission is absorbed by the ionized gas, and only 
the blueshifted OH peak will be present (\cite{rodriguez85}). 
Then a very young PNe will have a compact ionized inner part 
while the outer part of the envelope remains neutral, possibly with OH maser 
emission. It is possible to appreciate in several young OH-PNe, with known LSR 
velocities, that the blueshifted OH maser components are the dominant ones 
(\cite{Shepherd90}). 
For example, for 
NGC~6302 the ionized gas has a V$_{LSR}$ (He 121$\alpha$)= $-$29 km~s$^{-1}$ 
(\cite{Gomez87} ), while the strong OH 1612 MHz 
maser emission is blueshifted, arising from a velocity of $-$40 km~s$^{-1}$ 
(\cite{Payne88} ).
Several surveys of OH maser emission have being carried out toward PNe 
(e.g. \cite{Johansson77}; \cite{Casw81}; \cite{Payne88}; \cite{bowKnapp89}; 
\cite{Zijlstra89}; \cite{Sevenster01}) showing that it is 
uncommon to observe OH maser emission in PNe, with the detection of very
few objects (see Table~\ref{tab:young}). Then, if it is difficult to detect 
OH maser emission, the detection of H$_2$O masers in PNe is expected to be 
more rare, however our
group have confirmed the detection of two H$_2$O-PNe (\cite{Miranda01}; \cite{deGre04}) 
and another two H$_2$O-PNe have been
recently detected with the Robledo de Chavela Antenna (\cite{Suarez07}). 
In Table~\ref{tab:young}
we list the detections of H$_2$O maser emission in PNe.

\section{K~3-35 and IRAS~17347-3139}\label{sec:sources}

The first confirmed cases of PNe where H$_2$O maser emission is present
and coexistent with OH maser and radio continuum emission are: K~3-35 
(\cite{Miranda01}) and IRAS~17347-3139 (\cite{deGre04}). Both H$_2$O-PNe
are emission line nebulae that
exhibit bipolar morphologies at radio wavelengths (K~3-35: 
\cite{Aaquist93}; \cite{Miranda01}, IRAS~17347: \cite{Tafoya07b}) and 
the H$_2$O masers are
arising from the core of the sources in a torus-like structure 
(see figure~\ref{fig:torus}).  In the 
particular case of K~3-35, there
is also H$_2$O maser emission arising from the tips of the bipolar lobes at
5000 AU from the central star (\cite{Miranda01}; \cite{Gomez03}). 
Bipolar morphologies in these H$_2$O-PNe are also appreciated from optical/IR images
(\cite{SanchezC06}).
The systemic LSR velocities for these two H$_2$O-PNe, obtained from CO
observations, are 24$\pm$3 km~s$^{-1}$ 
for K~3-35 (\cite{Tafoya07a}) and 
$-$55 km~s$^{-1}$ for IRAS~17347-3139 
(\cite{Uscanga07}). The LSR velocities of the H$_2$O maser components in these
PNe appear blueshifted, $<$24 km~s$^{-1}$ for K~3-35 
(\cite{Miranda01}; \cite{Gomez03}; \cite{Tafoya07a}) and 
$\geq -$65 km~s$^{-1}$ for IRAS~17347-3139 
(\cite{deGre04}; \cite{Uscanga07}). 
In general, young PNe have a 
very high molecular to ionized gas mass ratio 
(\cite{Huggins96}; \cite{Josselin03}).
In the case of K~3-35, the ratio of molecular to ionized mass is $\sim$2 and 
the recent detection of HCO$^+$ emission suggests the presence of very high 
density molecular gas 
($\sim$10$^5$~cm$^{-3}$) that could be protecting the H$_2$O molecules from the 
dissociating radiation of the central star
(\cite{Tafoya07a}). The bipolar morphology,
and the presence of large amounts of neutral molecular gas not only suggest that
these objects are young, but also that they have followed a similar
evolutionary track corresponding to massive progenitors.

\begin{figure}
\includegraphics[height=3in,width=4in,angle=0]{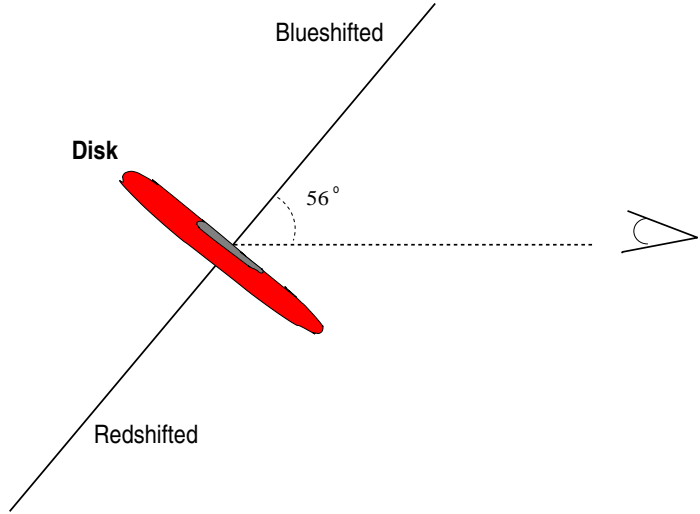}
  \caption{Schematic diagram for the inclinated disk model in K~3-35. 
}\label{fig:inclination}
\end{figure}
\section{Kinematics of the H$_2$O masers}\label{sec:sources}

Figure~\ref{fig:torus} shows radio continuum contour 
images and  the H$_2$O maser spot positions for K~3-35 and IRAS~17347-3139.
Modeling of the H$_2$O masers was made for a homogeneous disk-like structure
fitting each maser spot velocity with an expanding and rotating disk. For K~3-35
the model resulted in expanding and rotating velocity values of 1.1 km~s$^{-1}$ 
and 3.6 km~s$^{-1}$, respectively. The inclination angle of the disk, between 
the line-of-sight and the major axis of the outflow, was also derived (56$^\circ$) 
(see figure~\ref{fig:inclination}), which is in agreement with having the north lobe
of the outflow blueshifted and the south redshifted. The estimated radius of
the ring is 0${\rlap.}^{\prime\prime}$023 ($\sim$ 100 AU).  
For IRAS~17347-3139 it was more difficult to do the
modeling since the H$_2$O masers are interpreted as tracing only one edge of the 
equatorial disk.
However, in both cases the orientation of the major axis of the rings appears
perpendicular to the corresponding outflows, supporting a disk model interpretation 
(\cite{Uscanga07}).

\begin{figure}
\includegraphics[height=3in,width=5in,angle=0]{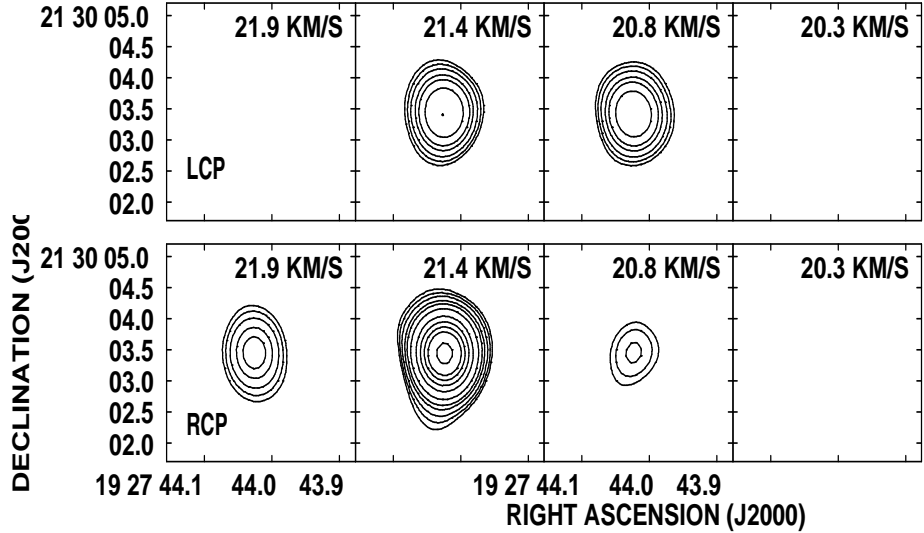}
  \caption{Contour images of OH 1720 MHz emission from K~3-35. Top panels are
for the left circular plarization (LCP) and the bottom panels are for the right
circular polarization (RCP) maser emission. The LSR velocities are indicated in the top
right-hand corner of each image. Contours are 5, 7, 10, 15, 20, 30, 50, 70, 90, 110
and 150 times 8 and 9 mJy~beam$^{-1}$, the $rms$ noise of the LCP and RCP images, respectively.
 }
\label{fig:magnetic}
\end{figure}

\section{Equatorial magnetic field in K~3-35}\label{sec:magnetic}

The detection of magnetic fields in PNe is a key parameter to understand the
generation of jets and bipolar structures in PNe (see Vlemings; these proceedings).
K~3-35 is one of the few PNe where magnetic fields have been detected (\cite{Gomez05}). 
In particular, the OH 1665 MHz maser emission toward K~3-35, was 
found to be very highly polarized ($>$50\%), suggesting
the presence of a magnetic field (\cite{Miranda01}). Since the 1665 MHz maser 
spots seems to be tracing an
elongated structure perpendicular to the bipolar outflow, it was proposed that they 
could be tracing an equatorial magnetic field. New VLA observations of the four
OH maser transitions confirmed the strong circular polarization of the OH 1665 MHz
emission and assuming that at least one Zeeman pair is present, we derived a magnetic
field in the line-of-sight of 0.14 mG at a radius of 250 AU. In addition, 
the OH 1720 MHz transition was 
clearly detected. A single OH 1720 MHz feature was previously reported by Te 
\cite{Telintel91} toward K~3-35. Our VLA OH observations confirmed the presence
of a single feature in the spectrum of the 1720 MHz and also proved that it is
associated with the H$_2$O-PNe, since the peak position of the maser feature coincides
with the peak of the radio continuum emission (\cite{Gomez05}). It is the only
one PNe where the OH 1720 MHz maser transition has been detected. 
A contour image of the OH 1720 MHz from K~3-35 in
its left and right circular polarizations is shown in figure~\ref{fig:magnetic}. 
Assuming that a Zeeman pair is present,
a magnetic field of 0.8 mG was derived at a radius of 120 AU.   
The strength of the magnetic field in K~3-35 is in agreement with the values
derived toward evolved stars.
\section{Conclusions}\label{sec:concl}

The presence of H$_2$O and OH maser emission in PNe is not common, and
their detection suggest that these objects are in a very early phase
of their evolution. In particular the presence of H$_2$O maser emission has
been confirmed only in two PNe (K~3-35 and IRAS~17347-3139), and from a
recent H$_2$O survey, made with the Robledo de Chavela Antenna, two more PNe 
seem to be detected. 
The H$_2$O maser positions in  K~3-35 and IRAS~17347-3139 show kinematics consistent
with masers located in rings. From modeling, it is derived that in K~3-35 there
is a rotating and expanding disk perpendicular to the bipolar outflow.
There is evidence of magnetic fields in K~3-35 that could help to constrain the
models for jets and bipolar morphologies in PNe. 

\begin{acknowledgments}
YG acknowledge financial support from PAPIIT-UNAM (IN100407)
and CONACyT, M\'exico. 
\end{acknowledgments}

\end{document}